\documentclass[prl,twocolumn,floatfix,showpacs]{revtex4}
\usepackage{graphicx}
\usepackage{amssymb,amsfonts,amsmath}
\usepackage{color}

\newcommand{\nc}{\newcommand}
\nc{\rnc}{\renewcommand}
\nc{\nn}{\nonumber}

\nc{\g}{\gamma}
\nc{\om}{\omega}
\rnc{\b}{\beta}
\rnc{\th}{\theta}

\rnc{\[}{\left[}
\rnc{\]}{\right]}
\rnc{\(}{\left(}
\rnc{\)}{\right)}

\newcommand{\bra}{\langle}
\newcommand{\ket}{\rangle}
\nc{\vac}{|0\ket}
\nc{\vvac}{\bra0|}

\nc{\cd}{\cdots}
\nc{\sm}[2]{\sum_{#1=1}^{#2}}

\nc{\red}{\textcolor{red}}
\nc{\sred}[1]{\textcolor{red}{\sout{#1}}}

\nc\hp{\hat{\psi}}
\nc\hpd{\hat{\psi}^\dagger}

\rnc{\[}{\left[}
\rnc{\]}{\right]}
\rnc{\(}{\left(}
\rnc{\)}{\right)}
\nc{\lt}{\left\{}
\nc{\rt}{\right\}}
\nc{\sg}{\sigma}
\nc{\lam}{\lambda}
\nc{\eps}{\epsilon}
\rnc{\a}{\alpha}
\rnc{\b}{\beta}
\nc{\vp}{\varphi}
\nc{\kp}{\kappa}
\rnc{\i}{{\rm i}}
\rnc{\d}{{\rm d}}

\begin{document}
\title{Finite-size scaling behavior of Bose-Einstein condensation 
in the 1D Bose gas 
}

\author{Jun Sato}
\author{Eriko Kaminishi}
\author{Tetsuo Deguchi}
\affiliation{Department of Physics, 
Graduate School of Humanities and Sciences, 
Ochanomizu University, 2-1-1 Ohtsuka, Bunkyo-ku, Tokyo 112-8610, Japan}

\date{\today}

\begin{abstract} 
Through exact numerical solutions we show Bose-Einstein condensation (BEC) 
for the one-dimensional (1D) bosons with repulsive short-range interactions 
at zero temperature by taking a particular large size limit. 
Following the Penrose-Onsager criterion of BEC, 
we define condensate fraction by the fraction of the largest eigenvalue 
of the one-particle reduced density matrix. 
We show the finite-size scaling behavior such that condensate fraction is given by a scaling function of one-variable: interaction parameter multiplied by a power of particle number. Condensate fraction is nonzero and constant for any large value of particle number or system size, if the interaction parameter is proportional to the negative power of particle number.  
Here the interaction parameter is defined by the coupling constant of the delta-function potentials devided by the density. 
With the scaling behavior we derive various themodynamic limits where condensate fraction is constant for any large system size; for instance, 
it is the case even in the system of a finite particle number.  
\end{abstract} 
\maketitle
The experimental realization of trapped atomic gases in one dimension 
has provided a new motivation for the study of 
strong correlations in fundamental quantum mechanical systems 
of interacting particles \cite{Ketterle,Esslinger,Kinoshita}. 
In one-dimensional (1D) systems quantum fluctuations 
play  a key role and often give subtle and nontrivial effects. 
It is known that Bose-Einstein condenstation (BEC) occurs even 
for bosons with repulsive interactions due to the quantum statistical effect 
among identical particles \cite{Leggett-book}. 
In fact, the existence of BEC has been proven rigorously 
for interactiong bosons confined in dimensions greater than one 
\cite{Lieb}. In 1D case there is no BEC for  
bosons with repulsive interactions due to strong quantum fluctuations 
if we take the standard thermodynamic limit with fixed coupling constant 
\cite{Pitaevskii}.  
On the other hand, if the coupling constant is very weak, 
we may expect that even the 1D bosons with a  
 large but finite number of particles 
undergo a quasi-condensation in which 
 ``a macroscopic number of particles occupy 
a single one-particle state'' \cite{Leggett-book}.

However, it has not been shown explicitly how such a quasi-condensation  
 occurs in interacting bosons in one dimension. 
Furthermore, it is nontrivial to expect it  
for the 1D Bose gas that is solvable by the Bethe ansatz. 
No pair of particles can have the same quasi-momentum 
in common for a Bethe-ansatz solution. 
Here we call the 1D system of bosons interacting 
with repulsive delta-function potentials  
the 1D Bose gas. For the impenetrable 1D Bose gas where  
the coupling constant is taken to infinity, 
condensate fractions are analytically and 
numerically studied \cite{Forrester}, 
while in the weak coupling case 
it is nontrivial to evalaute the fractions in the 1D Bose gas.

In the present Letter we show the finite-size scaling behavior 
of condensate fraction in the 1D Bose gas 
with repulsive interactions at zero temperature,   
and derive BEC by controlling the thermodynamic limit. 
We show that if the coupling constant decreases as a power of the system size,  condensate fraction does not vanish and remains constant 
when we send the system size to infinity 
or to a very large value with fixed density. 
If condensate fraction is nonzero for a large number of particles, 
we call it BEC according to the Penrose-Onsager criterion.

The scaling behavior of BEC in the 1D Bose gas is 
fundamental when we specify the thermodynamic limit, 
where we send particle number $N$ or system size $L$ to infinity 
or very large values. We define interaction parameter $\gamma$ 
by $\gamma=c/n$ with coupling constant $c$ in the delta-function potentials 
and density $n=N/L$. We show that if $\gamma$ is given by 
a negative power of $N$, i.e. $\gamma=A/N^{\eta}$, 
condensate fraction $n_0$ is nonzero and constant 
for any large value of $L$ or $N$. We also show that  
exponent $\eta$ and amplitude $A$ are independent of density $n$, 
and evaluate them as functions of $n_0$. 
Condensate fraction $n_0$ is thus given by 
a scaling function of variable $\gamma N^{\eta}$.  
If the condensate fraction of a quantum state with large $N$ 
is nonzero in the 1D Bose gas, 
we suggest that the classical mean-field approximation 
such as the Gross-Pitaevskii (GP) equation is valid 
for the state \cite{SKKD1}.  
Furthermore, we show that the 1D Bose gas of 
a finite particle number may 
have the same condensate fraction for any large $L$.

Let us review the definition of BEC 
through the one-particle reduced density matrix 
for a quantum system \cite{Leggett-book,PO}. 
 We assume that the number of particles $N$ is very large but finite.  
At zero temperature, the density matrix is given 
by $\hat{\rho}=|\lam\ket\bra \lam |$, 
where $|\lam\ket$ denotes the ground state of the quantum system. 
We define the one-particle reduced density matrix by 
the partial trace of the density matrix with respect to other 
degrees of freedom: $\hat{\rho}_1=N\text{tr}_{23\cd N}\hat{\rho}$. 
This matrix is positive definite and hence it is diagonalized as 
\begin{align}
\hat{\rho}_1=N_0|\Psi_0\ket\bra\Psi_0|+N_1|\Psi_1\ket\bra\Psi_1|
+\cd. 
\end{align}
Here we put eigenvalues $N_j$ in descending order:   
$N_0 \geq N_1 \geq N_2 \geq \cd>0$.  The sum of all the eigenvalues is given by the number of particles: $\sum_j N_j=N$. 
Here we recall $\text{tr}_1\hat{\rho}_1=N$ due to the normalization: 
$\text{tr}_{123\cd N} \hat{\rho} =1$.  
Let us denote by $n_0$ the ratio of 
the largest eigenvalue $N_0$ to  particle number $N$: 
\begin{align} 
n_0:= N_0/N .  
\end{align}
The criterion of BEC due to Penrose and Onsager \cite{PO} is given 
as follows: If the largest eigenvalue $N_0$ is of order $N$, i.e. 
 the ratio $n_0$ is nonzero and finite for large $N$,  
then we say that the system exhibits BEC, and  
we call $n_0$ the condensate fraction.  Here  
we also define fractions $n_j$ by $n_j=N_j/N$ for $j=1, 2, \ldots$.  

%
We now consider the Hamiltonian of the 1D Bose gas, which we call 
the Lieb-Liniger model (LL model) \cite{Lieb-Liniger}: 
\begin{align}
{\cal H}_{\text{LL}} 
= - \sum_{j=1}^{N} {\frac {\partial^2} {\partial x_j^2}}
+ 2c \sum_{j < k}^{N} \delta(x_j-x_k) . 
\end{align}
We assume the periodic boundary conditions 
of the system size $L$ on the wavefunctions. 
We employ a system of units with $2m=\hbar =1$, 
where $m$ is the mass of the particle. 
We consider the repulsive interaction: $ c > 0 $, hereafter.

In the thermodynamic limit the LL model is 
characterized by the single parameter $\gamma=c/n$, 
where $n=N/L$ is the density of particle number $N$.  
We fix the particle-number density 
as $n=1$ throughout the Letter, and change 
coupling constant $c$ so that 
we have different values of $\gamma$.  

In the LL model, the Bethe ansatz offers an exact eigenstate 
with an exact energy eigenvalue 
for a given set of quasi-momenta 
$k_1, k_2, \ldots, k_N$ satisfying 
the Bethe ansatz equations (BAE) 
for $j=1, 2, \ldots, N$: 
\begin{align}
 k_j L = 2 \pi I_j - 2 \sum_{\ell \ne j}^{N} 
\arctan \left({\frac {k_j - k_{\ell}} c } \right) . 
\label{BAE} 
\end{align}
Here $I_j$'s are integers for odd $N$ and half-odd integers for even $N$. 
We call them the Bethe quantum numbers. 
The total momentum $P$ and energy eigenvalue $E$ are written 
in terms of the quasi-momenta as 
\begin{align}
P=\sm{j}{N}k_j=\frac {2 \pi} L \sum_{j=1}^{N} I_j, \quad E=\sm{j}{N}k_j^2. 
\end{align}
If we specify a set of Bethe quantum numbers 
$I_1<\cdots<I_N$, BAE \eqref{BAE} have 
a unique real solution $k_1 < \cdots < k_N$ \cite{Korepin}. 
In particular, the sequence of the Bethe quantum numbers 
of the ground state is given by 
$I_j=-(N+1)/2+j$ for integers $j$ with $1 \leq j \leq N$. 
The Bethe quantum numbers for low lying excitations are 
systematically derived by putting holes or particles in the 
perfectly regular ground-state sequence.

The matrix element of the one-particle reduced density matrix, 
$\rho_1(x,y):=\bra x|\hat{\rho}_1|y\ket$, 
 for a quantum system is expressed 
as a correlation function in the ground state 
$| \lambda \rangle$: 
\begin{align}
\rho_1(x,y) = \bra \lam |\hat{\psi}^\dagger(y)\hat{\psi}(x)| \lam \ket. 
\end{align} 
In the LL model we can numerically 
evaluate the correlation function 
by the form factor expansion. 
Inserting the complete system of eigenstates, 
$\sum_{\mu} |\mu \rangle \langle \mu|$, we have 
\begin{align}
\rho_1(x,y)
=\sum_\mu
e^{\i(P_\mu-P_\lam)(y-x)}
| \bra \mu |\hat{\psi}(0)| \lam \ket |^2, 
\label{eq:sum-ff}
\end{align}
where $P_\mu$ denote the momentum eigenvalues of eigenstates $|\mu\ket$. 
Each form factor in the sum (\ref{eq:sum-ff}) is expressed as a  
product of determinants by making use of the determinant formula 
for the norms of Bethe eigenstates \cite{GK} and 
that for the form factors of the field operator 
\cite{Slavnov, Kojima, Caux-Calabrese-Slavnov2007}: 
\begin{align}
&\bra \mu|\hp(0)|\lam\ket
=(-1)^{N(N+1)/2+1}
\nn\\&\times
\(
\prod^{N-1}_{j=1}
\prod^N_{\ell=1}
\frac{1}{k'_j-k_\ell}\) 
\( 
\prod^N_{j>\ell}k_{j,\ell}\sqrt{k_{j,\ell}^2+c^2} 
\)
\nn\\&\times
\( 
\prod^{N-1}_{j>\ell}\frac{k'_{j,\ell}}{\sqrt{(k'_{j,\ell})^2+c^2}} 
\)
\frac{\det U(k,k')}{\sqrt{\det G(k)\det G(k')}} \, , 
\label{eq:Slavnov}
\end{align}
where the quasi-momenta $\{k_1,\cdots,k_N\}$ and $\{k'_1,\cdots,k'_{N-1}\}$ 
give the eigenstates $|\lam\ket$ and $|\mu\ket$, respectively. 
Here we have employed the abbreviated symbols 
$k_{j,\ell}:=k_j-k_\ell$ and $k'_{j,\ell}:=k'_j-k'_\ell$. 
The matrix $G(k)$ is the Gaudin matrix, whose $(j,\ell)$th element is 
$G(k)_{j,\ell}=\delta_{j,\ell}\[L+\sum_{m=1}^NK(k_{j,m})\]-K(k_{j,\ell})$
for $j, \ell=1,2,\cdots,N$, where the kernel $K(k)$ 
is defined by $K(k)=2c/(k^2+c^2)$. 
The matrix elements of the $(N-1)$ by $(N-1)$ matrix $U(k,k')$ are given by 
\cite{Caux-Calabrese-Slavnov2007, Slavnov, Kojima, GK} 
\begin{align}
U(k,k')_{j,\ell}&=2\delta_{j\ell}\text{Im}\[
\frac{\prod^{N-1}_{a=1}(k'_a-k_j + ic)}{\prod^N_{a=1}(k_a-k_j + ic)}\]
 \nn\\&
+\frac{\prod^{N-1}_{a=1}(k'_a-k_j)}{\prod^N_{a\neq j}(k_a-k_j)}
\(K(k_{j,\ell})-K(k_{N,\ell})\) . 
\label{eq:matrixU}
\end{align}

\begin{table} 
\begin{center} \begin{tabular}{cccc}
$c$  &    0.01   &  1 &  100 \\
\hline 
1p1h   & 0.999984   & 0.971538 & 0.693620  \\
2p2h     & $1.59454 \times 10^{-5}$   & 0.0280102 & 0.289056   \\
$n_{\rm sat}$     & 1.00000    & 0.999548 & 0.982676 \\
\hline \end{tabular}
\caption{Fraction $n_{\rm sat}$ of the reduced density operator 
at the origin, $\rho_1(0, 0)$, to the density $n$, 
evaluated by taking the sum over a large number of eigenstates 
$|\mu \rangle$ with 
one particle and one hole (1p1h) or with two particles 
and two holes (2p2h) for $N=L=50$ ($n=1$):   
$n_{\rm sat} = 
\left(  \sum_{\mu}^{1p1h} + \sum_{\mu}^{2p2h}  \right) 
| \bra \mu |\hat{\psi}(0)| \lam \ket |^2 /n . $}
\end{center} \end{table}

Numerically we  calculate correlation function (\ref{eq:sum-ff}) 
by taking the sum over a large number of eigenstates 
with one particle and one hole (1p1h) and 
those with two particles and two holes (2p2h). 
In order to confirm the validity of the restricted sum, 
we have estimated the ratio 
of the one-particle reduced density operator at the origin 
to density $n$,  $\rho_1(0, 0)/n$,  
through the form factor expansion (\ref{eq:sum-ff})  
for the excitations with 1p1h or 2p2h.  
We denote it by $n_{\rm sat}$. 
The estimates of $n_{\rm sat}$ are listed in Table 1.  
The graph of $n_{\rm sat}$ approaches 1 for small coupling constant $c$, 
while it is larger than 0.98 for any value of $c$ in the case of $N=50$.

\begin{figure}[t]
\includegraphics[width=0.40\textwidth]{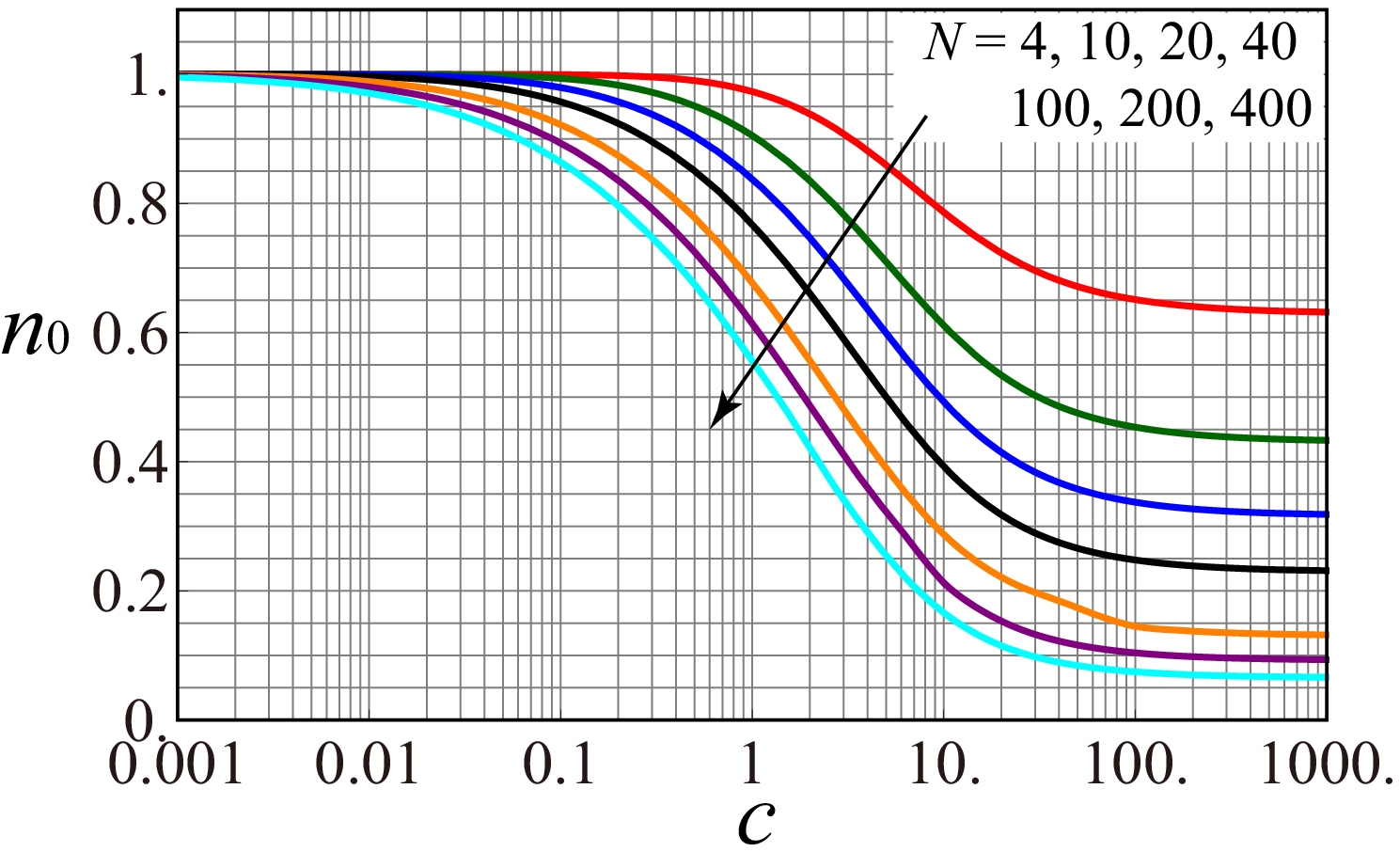}
\includegraphics[width=0.40\textwidth]{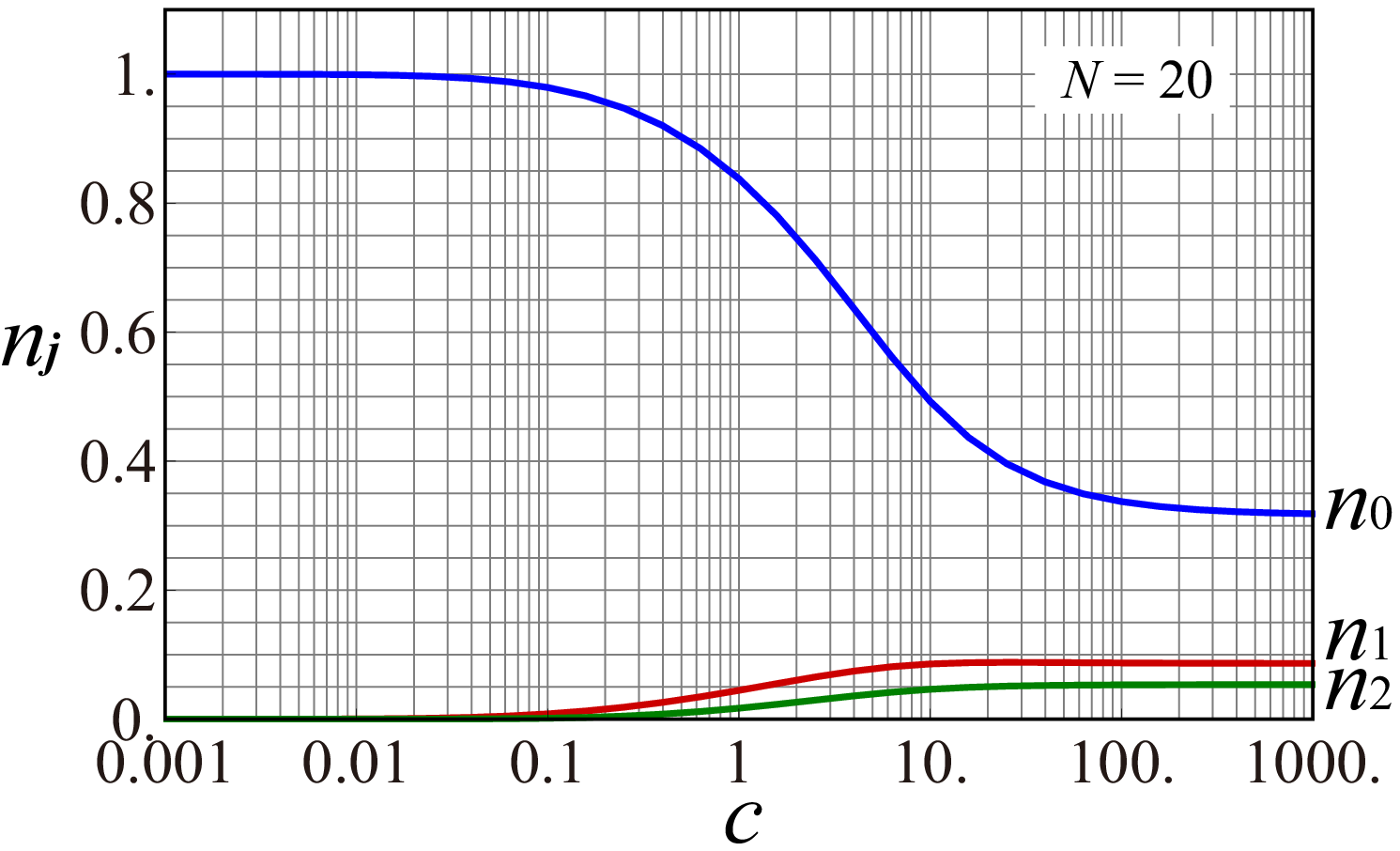}
\caption{(Color online) 
Dependence of fractions $n_j$ on coupling constant $c$. 
In the upper panel: condensate fraction $n_0$ is plotted against  
coupling constant $c$ 
for $N=4$, 10, 20, 40, 100, 200 and 400, 
from the top to the bottom, in red, green, blue, black, orange, purple 
and cyan lines, respectively. 
In the lower panel:  condensate fraction 
$n_0$,  fractions $n_1$ and $n_2$ are shown against $c$  
from the top to the bottom in blue, red and green lines, respectively, 
for $N=20$. We recall $n=N/L=1$. }
\label{c-cf}
\end{figure}

 For the LL model, 
the eigenfunctions of the one-particle reduced density matrix 
are given by plane waves for any nonzero and finite 
value of $c$. It is a consequence of the translational invariance of 
the Hamiltonian of the LL model. We thus have  
\begin{align}
\rho_1(x,y)
&=\frac{N_0}L+\sum_{j=1}^{\infty} \frac{2N_j}L \cos\[2\pi j (x-y)/L\] \, . 
\end{align} 
\par \noindent 
The eigenvalues of the one-particle reduced density matrix, 
$N_j$, are expressed in terms of the form factor expansion.  
We consider the sum over all the form factors between the ground state, 
$| \lambda \rangle$, and such eigenstates, $| \mu \rangle$,  
that have a given momentum $P_j$ as 
\begin{align}
N_j
=L\sum_{\mu: P_\mu=P_j}
| \bra \mu |\hat{\psi}(0)| \lam \ket |^2 \, .  \label{eq:Nj}
\end{align}
In the LL model we have $P_j:=(2\pi/L)j$.

Solving the Bethe ansatz equations for a large number of eigenstates 
we observe numerically that eigenvalues $N_j$ are given in decreasing order with respect to integer $j$: $N_0>N_1>N_2>\cd$. 
It thus follows that condensate fraction 
which corresponds to the largest eigenvalue of 
the one-particle reduced density matrix ${\hat \rho}_1$ 
is indeed given by $n_0=N_0/N$, 
where $N_0$ has been defined by sum (\ref{eq:Nj}) 
over all eigenstates with zero momentum.

The estimates of condensate fraction $n_0$ are plotted against 
coupling constant $c$ in the upper panel of Fig. \ref{c-cf} 
over a wide range of $c$ such as from $c=10^{-3}$ to $c=10^3$ 
for different values of particle number $N$ 
such as $N=4$, 10, \ldots, 400. 
 For each $N$, condensate fraction $n_0$ becomes 1.0 for small $c$ 
such as $c < 0.01$, while it decreases with respect to $c$ and  
approaches an asymptotic value in the large $c$ region such as 
$c > 100$ or 1000.  
The asymptotic values depend on particle number $N$ 
for $N=4$, 10, \ldots, 400, and they are consistent 
with the numerical estimates of occupation numbers 
for the impenetrable 1D Bose gas (see eq. (56) of Ref. \cite{Forrester}).   
In the lower panel of Fig. \ref{c-cf}, 
we plot fractions $n_j$ for $j=0,1$ and 2 against coupling constant 
$c$ from $c=10^{-3}$ to $c=10^3$ with $N=20$. 
The asymptotic values of $n_j$ for large $c$  (i.e. $c=1000$) 
are consistent with the numerical estimates for the impenetrable 1D Bose gas 
(for $n_1$ and $n_2$, see eqs. (57) and (58) of Ref. \onlinecite{Forrester},  
respectively).

\begin{figure}[t]
\includegraphics[width=0.4 \textwidth]{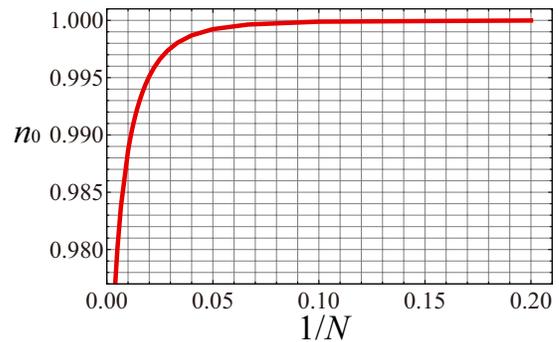} 
\caption{(Color online) Condensate fraction $n_0$ as a function of  
$1/N$ for $c=0.01$. Here $n=N/L=1.0$. 
}
\end{figure}

We observe that condensate fraction $n_0$ decreases 
as particle number $N$ increases when density $n=N/L$ is fixed.  
It is the case for $c < 0.1$  in the upper panel of Fig. 1.
Condensate fraction $n_0$ decreases as $N$ increases 
even for small $c$ such as $c=0.01$, as shown in Fig. 2. 
Thus, it is necessary for coupling constant $c$ to decrease 
with respect to $N$ so that condensate fraction $n_0$ remains 
constant as $N$ increases with fixed density $n$.

\begin{figure}[t]
\includegraphics[width=0.5\textwidth]{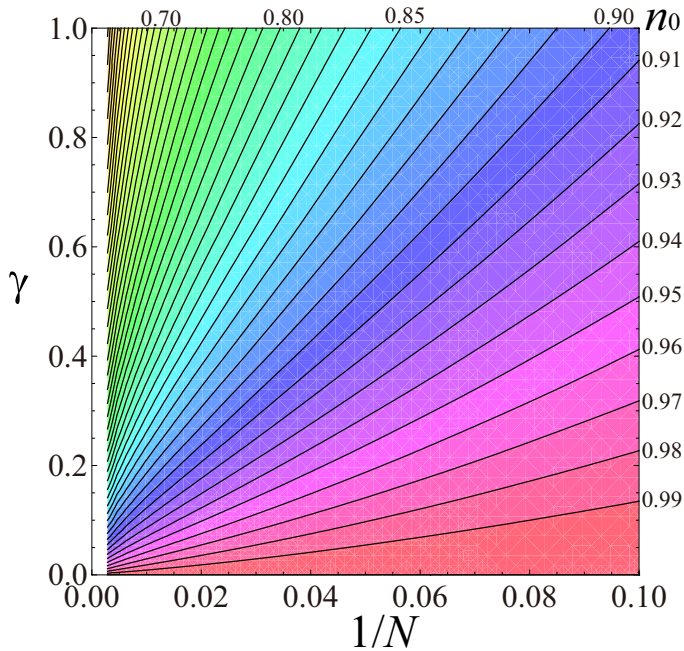}
\caption{(Color online) Contours of condensate fraction 
$n_0$ are plotted for various values of $n_0$ 
in the $\gamma$ versus $1/N$ plane.  
Each contour is approximated by (\ref{eq:scaling}): 
$\gamma$ as a function of $1/N$.   
}
\label{fig:countours}
\end{figure}

We now show the finite-size scaling of condensate fraction $n_0$. 
In Fig. \ref{fig:countours} 
each contour line gives the graph of interaction parameter  
$\gamma$ as a function of the inverse of particle number $N$ 
for a fixed value of condensate fraction $n_0$.  
They are plotted for various values of $n_0$ from $n_0=0.6$ to 0.99, 
and are obtained by solving the Bethe-ansatz equations numerically. 
For different values of density such as $n=1$, 2 and 5,   
we plot countour lines with fixed values of condensate fraction $n_0$ 
in the plane of interaction parameter $\gamma$ versus 
inverse particle number $1/N$. We observe that the contours 
with the same condensate fraction $n_0$ for the different densities  
coincide in the $\gamma$ versus $1/N$ plane 
and are well approximated by   
\begin{equation}
\gamma = A/N^{\eta} . \label{eq:scaling}
\end{equation}
Thus, condensate fraction $n_0$ is constant 
as particle number $N$ becomes very large 
if interaction parameter $\gamma$ is given by  
the power of particle number $N$ as in eq. (\ref{eq:scaling}).

Applying the finite-size scaling arguments, 
we suggest from eq. (\ref{eq:scaling}) that condensation fraction 
$n_0$ is given by a scaling fuction $\phi(\cdot)$ 
of a single variable $\gamma N^{\eta}$: $n_0=\phi(\gamma N^{\eta})$. 
Here we recall the coincidence of contours for the different values of density 
$n$ in Fig. \ref{fig:countours}.  
We thus observe that exponent $\eta$ and amplitude $A$ 
of eq. (\ref{eq:scaling}) are determimed only by condensate fraction $n_0$ 
and are independent of density $n$.  

\begin{figure}[t]
\includegraphics[width=0.35\textwidth]{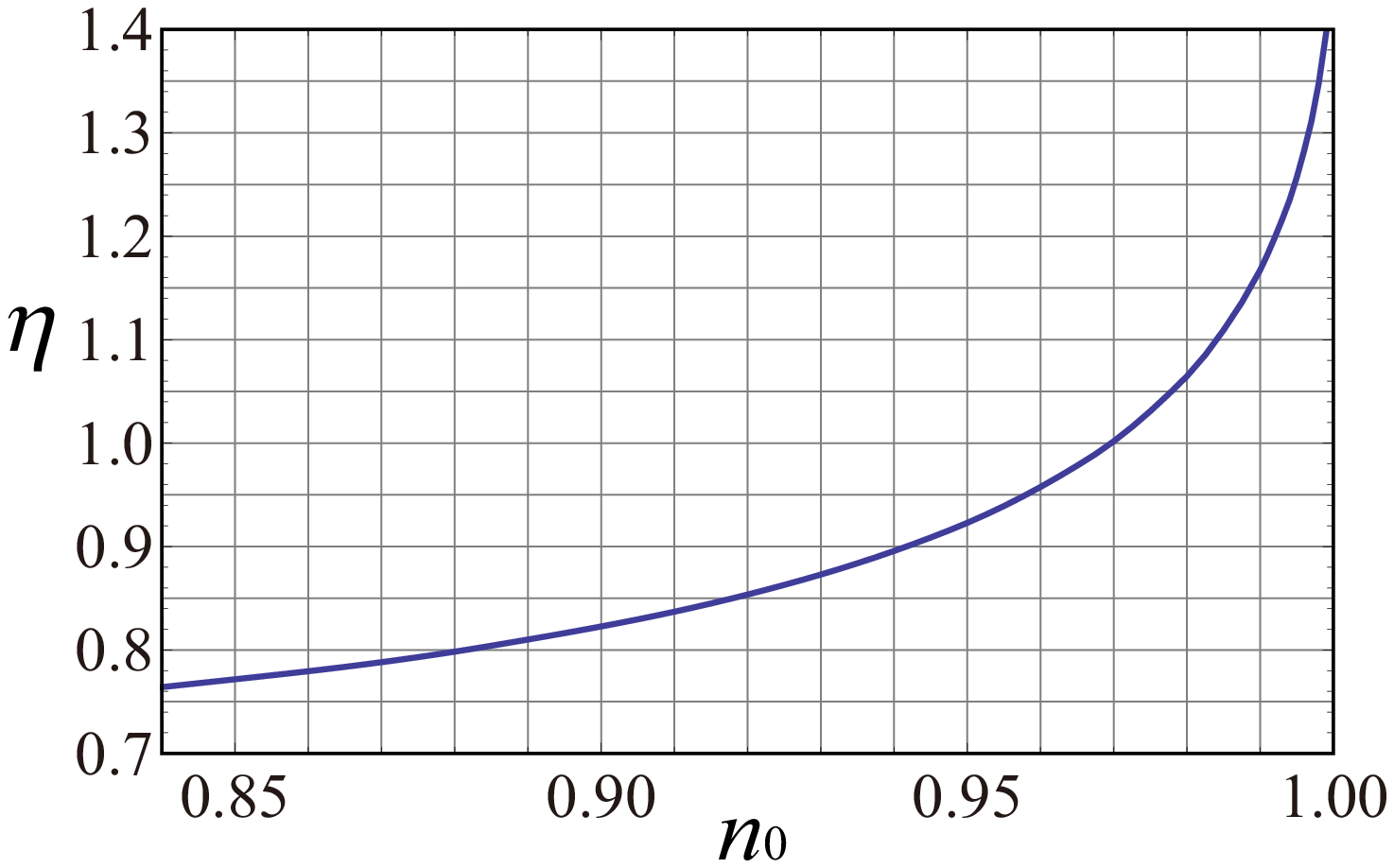}
\includegraphics[width=0.35\textwidth]{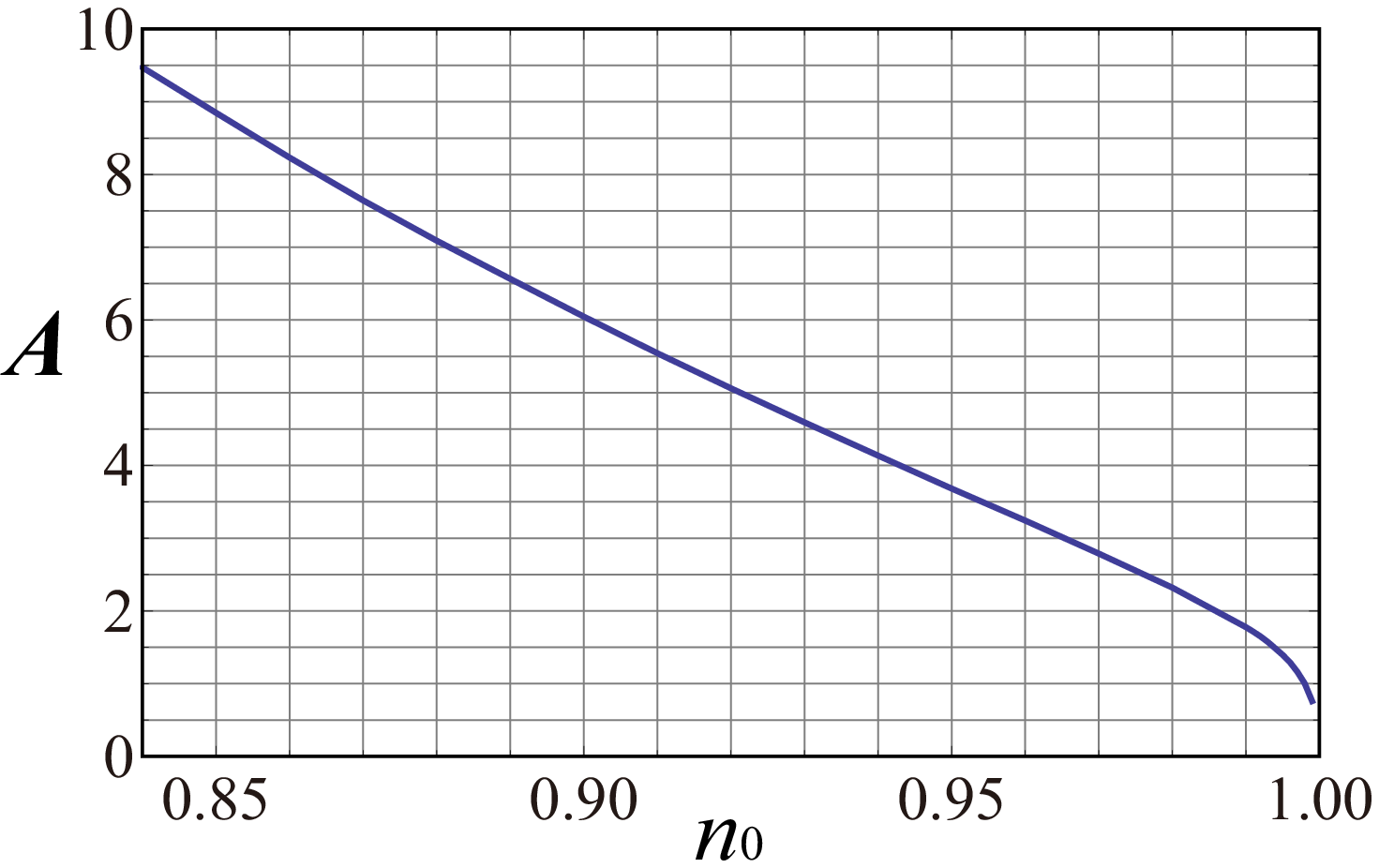}
\caption{(Color online) 
Exponent $\eta$ and amplitude $A$ 
as functions of condensate fraction $n_0$.   
}
\label{fig:eta-A}
\end{figure}

Let us consider amplitude $A$ as a function of $n_0$. 
We denote it by $A=f(n_0)$.   
Then, the scaling function $\phi(\cdot)$ is given by 
the inverse function: $n_0=f^{-1}(A)$.  
In Fig. \ref{fig:eta-A}, exponent $\eta$ increases with respect to $n_0$, and 
amplitude $A$ decreases monotonically with respect to $n_0$.

It follows from (\ref{eq:scaling}) that 
BEC does not occur in the 1D Bose gas if we fix 
parameter $\gamma$ and density $n$ as system size $L$ goes to infinity.  
However, if $\gamma$ is small enough 
so that it satisfies eq. (\ref{eq:scaling}) 
for a given value of condensate fraction $n_0$, 
the 1D Bose gas shows BEC from the viewpoint 
of the Penrose and Onsager criterion. 
We suggest that if condensate fraction $n_0$ of a quantum state 
is nonzero and finite for large $N$,  
the mean-field approximation is valid for the quantum state. 
For instance, there exist such quantum states 
that correspond to classical dark solitons of the GP equation \cite{SKKD1}, 
if parameter $\gamma$ is small enough so  
that it satisfies (\ref{eq:scaling}).

With the scaling behavior (\ref{eq:scaling}) 
we derive various ways of the thermodynamic limit 
such that condensate fraction $n_0$ is constant. 
For instance, we consider 
the case of a finite particle number, $N=N_{\rm f}$. 
Choosing a value of $n_0$, we determine $\gamma$ 
by eq. (\ref{eq:scaling}) as $\gamma= A(n_0)/N_{\rm f}^{\eta(n_0)}$.      
Then, the 1D Bose gas with $N=N_{\rm f}$ 
has the same condensate fraction $n_0$ for any large value of $L$ 
if coupling constant $c$ is given by $c=A(n_0) N_{\rm f}^{1- \eta}/L$. 
Let us set $\eta=1$ and $N_{\rm f}=10$, for simplicity. 
We have $n_0=0.97$ in Fig. \ref{fig:eta-A},    
and $\gamma = 0.3$ at $1/N=0.1$ in the contour of $n_0=0.97$ 
in Fig. \ref{fig:countours}. 
By assuming $n=1$, it corresponds 
to the case of $L=10$ and $c = 0.3$, and we have $A=c L = 3$, 
which is consistent with Fig. \ref{fig:eta-A}. 
Therefore, the 1D Bose gas with $N_{\rm f}=10$ 
has  $n_0 = 0.97$ for any large $L$ if $c$ is given by $c=0.3/L$.  
Moreover, we may consider other types of thermodnamic limits. 
When density $n$ is proportional to a power of $L$ as $L^{\alpha}$, 
condensate fraction 
$n_0$ is constant as $L$  goes to infinity if we set 
$c \propto L^{(1-\eta)(1+\alpha)-1}$.

The scaling law (\ref{eq:scaling}) and the estimates of condensate fraction 
in the present Letter should be useful for estimating conditions 
in experiments of trapped cold atomic gases in one dimension \cite{PS}. 
For instance, we suggest from Fig. 1 that 
BEC may appear in 1D systems with a small number of bosons 
such as $N=20$ or $40$ for $c=1$ or $10$.

In conclusion, we  exactly calculated the 
condensate fraction of the 1D Bose gas with repulsive interaction  
by the form factor expansion. 
We have shown the finite-size scaling behavior 
such that condensate fraction $n_0$ is given by a 
scaling function of interaction parameter $\gamma$ times 
some power of particle number $N$: $n_0=\phi(\gamma N^{\eta})$.  
Consequently, if parameter $\gamma$ decrease as $\gamma=A/N^{\eta}$, 
condensate fraction $n_0$ remains nonzero and constant 
as particle number $N$ becomes very large.   
By modifying the thermodynamic limit, 
the 1D Bose gas shows BEC from the viewpoint of 
the Penrose-Onsager criterion.

The authors thank R. Kanamoto for useful discussions. 
The present research is partially supported by Grant-in-Aid for Scientific Research No. 21710098 and No. 24540396. J.S. and E.K. is supported by JSPS.

\end{document}